\documentclass[conference]{IEEEtran}
\IEEEoverridecommandlockouts
\usepackage{cite}
\usepackage{amsmath,amssymb,amsfonts}
\usepackage{algorithmic}
\usepackage{graphicx}
\usepackage{textcomp}
\usepackage{xcolor} 

\usepackage{algorithm2e}
\usepackage{tabularx}
\usepackage{soul} 

\usepackage{makecell}
\usepackage{siunitx}
\usepackage{multirow}

\def\BibTeX{{\rm B\kern-.05em{\sc i\kern-.025em b}\kern-.08em
    T\kern-.1667em\lower.7ex\hbox{E}\kern-.125emX}}

\newcommand{\authorblockspace}{\hspace{1.5cm}}
\newcommand{\authorblockspacebis}{\hspace{-3.5cm}}

\begin{document}

\title{The Cost of Simplicity: \\
	How Reducing EEG Electrodes Affects Source Localization and BCI Accuracy \\
\thanks{This work is supported in part by STI 2030
	Major Projects, the National Key Research and Development Program of China and the Natural Science Foundation of China (Grant No. 2022ZD0208500).}
}

\author{\IEEEauthorblockN{Eva Guttmann-Flury }
\IEEEauthorblockA{
	\textit{Department of Micro-Nano Electronics and } \\
	\textit{the MoE Key Laboratory of Artificial Intelligence} \\
	\textit{Shanghai Jiao Tong University}\\
	Shanghai, China \\
	\textit{and visiting at CenBRAIN Neurotech,} \\
	\textit{School of Engineering, Westlake University}\\
	Hangzhou, China \\
	eva.guttmann.flury@gmail.com}
\and
\IEEEauthorblockN{Yanyan Wei }
\IEEEauthorblockA{
	\textit{Shanghai Key Laboratory of} \\ 
	\textit{Psychotic Disorders,}\\
	\textit{Shanghai Mental Health Center} \\
	\textit{Shanghai Jiao Tong University,} \\
	\textit{School of Medicine}\\
	Shanghai, China \\
	weiyanyan19860729@126.com}
\and
\IEEEauthorblockN{Shan Zhao }
\IEEEauthorblockA{
	\textit{School of Public Health} \\
	\textit{Shanghai Jiao Tong University,} \\
	\textit{School of Medicine}\\
	Shanghai, China \\
	shanzhao23@sjtu.edu.cn}
\and
\authorblockspace
\IEEEauthorblockN{Jian Zhao}
\IEEEauthorblockA{
	\authorblockspace
	\textit{Department of Micro-Nano} \\
	\authorblockspace
	\textit{Electronics and the MoE Key} \\
	\authorblockspace
	\textit{Laboratory of Artificial Intelligence} \\
	\authorblockspace
	\textit{Shanghai Jiao Tong University}\\
	\authorblockspace
	Shanghai, China \\
	\authorblockspace
	zhaojianycc@sjtu.edu.cn}
\and
\authorblockspacebis
\IEEEauthorblockN{Mohamad Sawan}
\IEEEauthorblockA{
	\authorblockspacebis
	\textit{CenBRAIN Neurotech,} \\
	\authorblockspacebis
	\textit{School of Engineering} \\
	\authorblockspacebis
	\textit{Westlake University}\\
	\authorblockspacebis
	Hangzhou, China \\
	\authorblockspacebis
	sawan@westlake.edu.cn}

}

\maketitle

\begin{abstract}
Electrode density optimization in electroencephalography (EEG)-based Brain-Computer Interfaces (BCIs) requires balancing practical usability against signal fidelity, particularly for source localization.
Reducing electrodes enhances portability but its effects on neural source reconstruction quality and source connectivity — treated as proxies to BCI performance — remain understudied. 
We address this gap through systematic evaluation of 62-, 32-, and 16-channel configurations using a fixed, fully automated processing pipeline applied to the well-characterized P300 potential. This approach's rationale is to minimize variability and bias inherent to EEG analysis by leveraging the P300's stimulus-locked reproducibility and pipeline standardization. Analyzing 63 sessions (31 subjects) from the Eye-BCI dataset with rigorous artifact correction and channel validation, we demonstrate: (1) Progressive degradation in source reconstruction quality with sparser configurations, including obscured deep neural generators and spatiotemporal distortions; (2) A novel $\sqrt{R_e}$ scaling law linking electrode reduction ratio ($R_e$) to localization accuracy -- a previously unquantified relationship to the best of our knowledge; (3) While reduced configurations preserve basic P300 topography and may suffice for communicative BCIs, higher-density channels are essential for reliable deep source reconstruction. Overall, this study establishes a first step towards quantitative benchmarks for electrode selection, with critical implications for clinical BCIs requiring anatomical precision in applications like neurodegenerative disease monitoring, where compromised spatial resolution could mask pathological signatures. Most importantly, the $\sqrt{R_e}$ scaling law may provide the first principled method to determine the minimal electrode density required based on acceptable error margins or expected effect sizes.

\end{abstract}


\section{Introduction}
Brain-Computer Interfaces (BCIs) establish direct communication between neural activity and external devices, supporting various assistive technologies. Data collected from BCI experiments also hold potential for applications in medical diagnostics, including the identification of biomarkers that could be relevant for neurological conditions (e.g., Alzheimer's disease). Electroencephalography (EEG), which measures electrical brain activity through scalp electrodes, remains the predominant non-invasive BCI modality due to its excellent temporal resolution, cost-effectiveness, and portability. However, EEG's limited spatial resolution and noise sensitivity present fundamental challenges for accurate neural source localization and reliable BCI performance \cite{Grover_2016, Chaudhary_2016, Hauk_2022}. 

\begin{figure}[!htp]
	\centering
	\includegraphics[width=8cm]{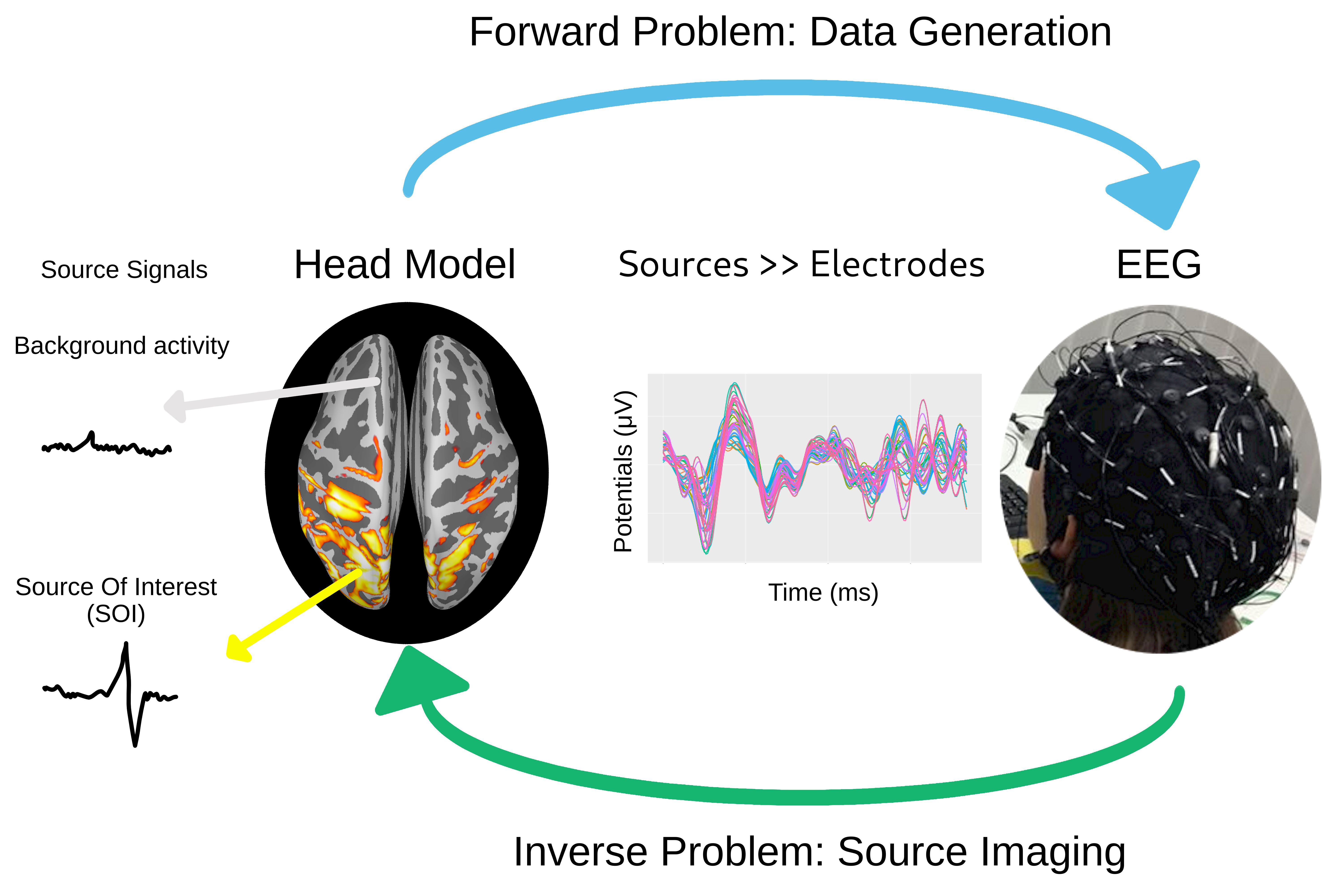}
	\caption
	[InvPb]
	{Overcoming the EEG inverse problem: When $10^2$ electrodes must map $10^{10}$ potential neural sources}
	\label{fig:InvPb}
\end{figure}

The core challenge of EEG source localization stems from the complex biophysics of electrical signal propagation through head tissues. Cortical pyramidal neuron activity generates electrical potentials that attenuate and disperse through intermediate tissues, creating distorted scalp measurements. The inverse problem, illustrated in Figure \ref{fig:InvPb},  involves reconstructing intracranial sources from these distorted signals, yet it is ill-posed, with infinite possible solutions. While anatomical constraints help restrict possible source configurations, current methods still struggle to distinguish adjacent cortical areas or resolve deep sources, which are critical for enhancing epilepsy treatment outcomes \cite{Grova_2008, Sohrabpour_2015}.

EEG source imaging accuracy faces three primary limitations. First, head model inaccuracies - especially in skull conductivity estimation - introduce systematic localization errors (10 -- 30 mm) due to individual anatomical variations \cite{AkalinAcar_2013, Unnwongse_2023}. Second, spatial sampling constraints limit resolution: while high-density arrays (128-256 channels) improve field topography reconstruction, standard 19-32 electrode setups may undersample critical spatial frequencies \cite{Brodbeck_2011, Eom_2022}. Third, signal contamination from physiological artifacts (muscle/ocular activity) and environmental noise disproportionately impacts estimation of high-frequency or deep sources \cite{TrujilloBarreto_2004, Michel_2019}.

Electrode density requirements remain controversial across studies due to methodological and application-specific differences. Exact Low Resolution Electromagnetic Tomography (eLORETA) performance shows clear dependence on channel count, with increasing electrode numbers improving both accuracy and active source detection \cite{Dattola_2020, Sohrabpour_2015}. These patterns, primarily established in epilepsy research using high-density arrays, raise the question of whether similar results can be generalized to BCI paradigms that require both spatial precision and real-time processing. This highlights the need for systematic evaluation of electrode reduction effects within standardized BCI frameworks that account for both algorithmic requirements and practical implementation constraints.

To address these gaps, this study utilizes eLORETA for its mathematically guaranteed source localization accuracy and inherent noise robustness \cite{PascualMarqui_2018}, providing a standardized framework to minimize methodological variability. We employ the well-characterized P300 potential as our signal of interest, leveraging its: (1) highly reproducible, stimulus-locked nature that reduces inter-subject variability; (2) distributed cortical-subcortical generators that comprehensively test spatial resolution; and (3) established clinical utility in cognitive assessment \cite{Hedges_2016}. The P300's spatiotemporal characteristics (P3a: 250 -- 280 ms at frontocentral; P3b: 300 -- 500 ms at parietal) enable systematic assessment of how electrode reduction affects both temporal precision and spatial accuracy.

The current investigation examines EEG electrode reduction effects by quantitatively comparing 62-, 32-, and 16-channel configurations through complementary metrics: (1) spatial localization accuracy and (2) source connectivity. The primary objectives are to: (i) establish the minimal electrode count that preserves clinically-relevant neural features, (ii) quantify how spatial resolution degrades with channel reduction, and (iii) provide evidence-based guidelines for optimizing BCI design. By employing this dual-metric approach, we aim to bridge the gap between theoretical source localization performance and practical BCI implementation constraints.

\section{Methods}

\subsection{Data Description}
The Eye-BCI multimodal dataset \cite{eye_bci_multi_dataset} provides 62-channel EEG recordings from 31 neurologically intact participants (age range: 22-57 years; 29 right-handed) during 2,520 P300 trials. This dataset offers three critical features for investigating electrode reduction effects: (1) comprehensive multimodal recordings including synchronized EOG, EMG, and eye-tracking for artifact correction \cite{GuttmannFlury_Dataset_2025}; (2) a sample size determined through Fitted Distribution Monte Carlo power analysis \cite{GuttmannFlury_APriori_2019} to ensure statistical sensitivity; and (3) fully documented experimental protocols supporting reproducibility, available at \cite{eye_bci_multi_dataset}.

\subsection{Signal Preprocessing Pipeline}

The EEG preprocessing begins with rigorous artifact detection and channel validation procedures. The Adaptive Blink and Correction De-drifting (ABCD) algorithm \cite{GuttmannFlury_BadChannel_2025} identifies blink events and computes electrode-specific Blink-Related Potentials (BRP) through averaging. Channel quality assessment is performed by comparing each electrode's BRP to the averaged BRP of its neighboring electrodes. Electrodes with abnormal BRP patterns, indicating potential hardware issues or poor contact, are automatically excluded to ensure data quality. This standardized procedure consistently excludes problematic electrodes across all recordings.

Following channel rejection, the signals undergo ABCD to eliminate drift and blink artifacts while preserving the original 1000 Hz sampling rate to enable millisecond-level temporal analysis of P300 components. Spatial filtering via a surface Laplacian transform \cite{Kayser_2015} enhances topographic resolution by emphasizing local cortical generators and attenuating volume-conducted interference. Subsequent temporal filtering employs a zero-phase 4th-order Butterworth bandpass (1-15 Hz) to selectively preserve P300-relevant frequencies, effectively suppressing both high-frequency noise and slow cortical potentials below 1 Hz. The preprocessing concludes with epoch segmentation from -200 ms to +700 ms relative to stimulus onset, preserving the complete P300 dynamics while optimizing signal quality for subsequent source analysis.

\subsection{Anatomical Registration and Electrode Configurations}
\begin{figure}[!htp]
	\centering
	\includegraphics[width=7cm]{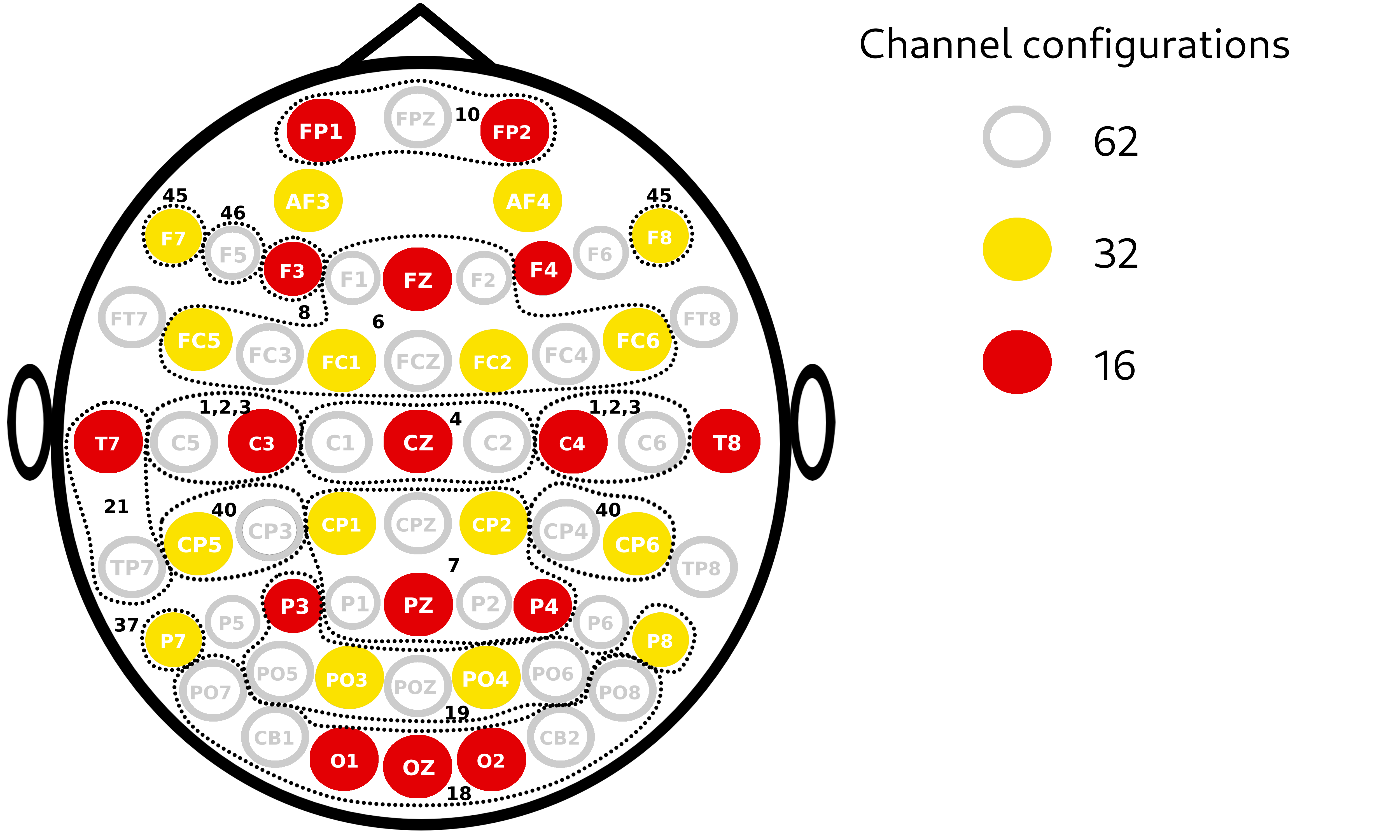}
	\caption
	[Electrode configurations and Brodmann area coverage]
	{Three EEG montages with their associated Brodmann areas.}
	\label{fig:DiffElec}
\end{figure}

Electrode positions follow standard international 10-5, 10-10, and 10-20 system configurations, which are the most widely adopted montages in both research and clinical EEG applications. The Koessler 3D anatomical atlas \cite{Koessler_2009} serves as the reference framework for mapping these standardized positions to Talairach coordinates and cortical surfaces. The most probable Brodmann area for each electrode is determined through population-level neuroanatomical analysis. Three representative configurations are evaluated: medium- (62 channels), low- (32 channels), and reduced-density (16 channels), as shown in Figure \ref{fig:DiffElec}. These configurations are selected to reflect common experimental and clinical practices while maintaining methodological consistency and allowing rigorous comparison of spatial sampling effects.

\subsection{Mathematical Approaches to Source Estimation}
The EEG inverse problem is severely underdetermined, with approximately \(10^9\) -- \(10^{10}\) potential neural sources vastly outnumbering typical \(10^1\) -- \(10^2\) recording electrodes \cite{HerculanoHouzel_2009}. The limited number of electrodes relative to the vast number of potential sources underscores the critical role of electrode count in source localization accuracy. Volume conduction effects also cause electrical currents to diffuse non-uniformly through cerebrospinal fluid, skull, and scalp tissues, blending contributions from distinct neural populations. These challenges are compounded by measurement noise and biological artifacts that disproportionately affect source estimation accuracy.

Two principal mathematical frameworks address these issues. Parametric methods assume focal activation patterns modeled by a small number of equivalent current dipoles (typically 1 -- 5 sources), making them suitable for evoked responses or epileptic spike localization but less appropriate for distributed cognitive processes. In contrast, distributed methods represent the solution space as a three-dimensional voxel grid (typically containing 2,000 -- 20,000 points), with each voxel containing a current dipole characterized by three moment components. This approach enables whole-brain reconstruction without requiring a priori assumptions about source locations \cite{Dattola_2020}.

The forward model mathematically relates neural sources to scalp measurements through the lead field matrix K \cite{PascualMarqui_2018}:
\begin{equation}
	\Phi = K J + c1
\end{equation}

 where $\Phi \in \mathbb{R}^{N_E \times 1}$ represents scalp potentials at $N_E$ electrodes, $K \in \mathbb{R}^{N_E \times (3 N_V)}$ is the lead field matrix for $N_V$ voxels, $J \in \mathbb{R}^{(3 N_V) \times 1}$ contains the current density moments, $c$ accounts for the physical property that electric potentials are determined up to an arbitrary constant, and $1 \in \mathbb{R}^{N_E \times 1}$ denotes a vector of ones. The inverse solution is obtained through the regularized linear transformation:
 \begin{equation}
 	\hat{J} = T \Phi
 \end{equation}
where $T$ is the generalized inverse of $K$. Tikhonov regularization balances noise suppression with fidelity; as noise and regularization vanish, the estimate converges to the minimum-norm, physiologically plausible solution \cite{Nemaire_2023}.


\subsection{Zero-Error Source Localization}
The exact Low Resolution Electromagnetic Tomography (eLORETA) method offers a three-dimensional reconstruction of neural electrical activity using the digitized brain model developed by the Brain Imaging Center of the Montreal Neurological Institute (MNI). This model constrains the solution space to cortical gray matter and hippocampal regions, focusing on areas of significant neural activity. eLORETA extends the weighted minimum norm estimation approach through a novel weighting scheme that simultaneously addresses depth compensation and spatial smoothing. The solution is obtained through a regularized linear transformation:

 \begin{equation}
 	\hat{J}_W = T_W \Phi
 \end{equation}
 where the pseudoinverse $T_W$ is defined as:
  \begin{equation}
 	T_W = W^{-1} K^T (K W^{-1} K^T + \alpha H)^{+}
 \end{equation}
 

The block-diagonal weight matrix $W \in \mathbb{R}^{(3 N_V) \times (3 N_V)}$ embeds Laplacian smoothing and depth weighting in each 3×3 voxel subblock, while the regularization parameter $\alpha \geq 0$ balances accuracy against noise and $H$ denotes the average-reference centering matrix; the superscript "+" signifies the Moore-Penrose pseudoinverse, which collapses to the standard inverse for non-singular systems.

The spatial Laplacian component of $W$ enforces the neurophysiological principle that neighboring cortical areas exhibit correlated neural activity, while the depth weighting compensates for the inherent bias toward superficial sources in EEG measurements. This combined approach yields three key advantages over conventional LORETA: (1) mathematically proven zero localization error for single test sources under noisy conditions, (2) standardized unit variance enabling cross-subject comparisons, and (3) improved depth sensitivity reducing localization error for deep sources from $12\,mm$ to $7\,mm$. These properties make eLORETA particularly suitable for investigating electrode reduction effects while maintaining robust performance across different channel configurations.

\subsection{Head Model Reconstruction}
The accuracy of EEG source localization fundamentally depends on the fidelity of the head model used for forward calculations. While subject-specific head models derived from individual Magnetic Resonance Imaging (MRI) scans and tissue conductivity measurements would provide optimal accuracy, such data were not available for the participants in this study. Instead, the analysis employs the fsaverage template from FreeSurfer, a well-validated average brain model constructed from high-resolution MRI scans of 40 healthy adults \cite{Fischl_1999}.

The adult template MRI (fsaverage) employs a sophisticated surface-based registration approach that preserves cortical topography while enabling cross-subject comparisons. This process involves three key steps: (1) individual cortical surfaces are mapped onto spheres using maximally isometric transformations, (2) surfaces are morphed into register with a canonical average surface using both folding patterns (sulcal/gyral alignment) and isometry-preserving forces, and \linebreak (3) a unified coordinate system is established for surface-based averaging across subjects.

For forward modeling, the MNE-Python implementation constructs a Boundary Element Model (BEM) using template-specific tissue conductivity values: scalp ($0.33\,S/m$), skull ($0.006\,S/m$), and brain ($0.33\,S/m$). The BEM comprises triangulated surfaces representing tissue boundaries, with source space defined using an icosahedral subdivision (ico5) yielding 10,242 sources per hemisphere. This configuration provides a balance between computational efficiency and spatial resolution, with a source spacing of $3.1\,mm$ and a surface area of $9.8\,mm^2$ per source \cite{Gramfort_2013}.

\section{Results}

To minimize variability from manual execution and parametric tuning, an identical, fully automated processing pipeline is applied to all three electrode configurations. This approach ensures that observed differences in source localization and connectivity metrics are due to electrode reduction alone.

\subsection{Neural Source Reconstruction}

The source reconstruction pipeline processes each epoch independently using the average head model. Prior to inverse solution computation, the leadfield matrix undergoes noise pre-whitening based on the noise covariance matrix derived from the pre-stimulus baseline period (-200 to 0 ms relative to stimulus onset). This period's covariance estimation can become numerically unstable with limited samples, potentially introducing spurious correlations between source amplitude estimates and sample size. To mitigate this, we apply regularization to the noise covariance matrix, which is particularly crucial for epochs with restricted temporal samples.

\begin{figure}[!htp]
	\centering
	\includegraphics[width=8cm]{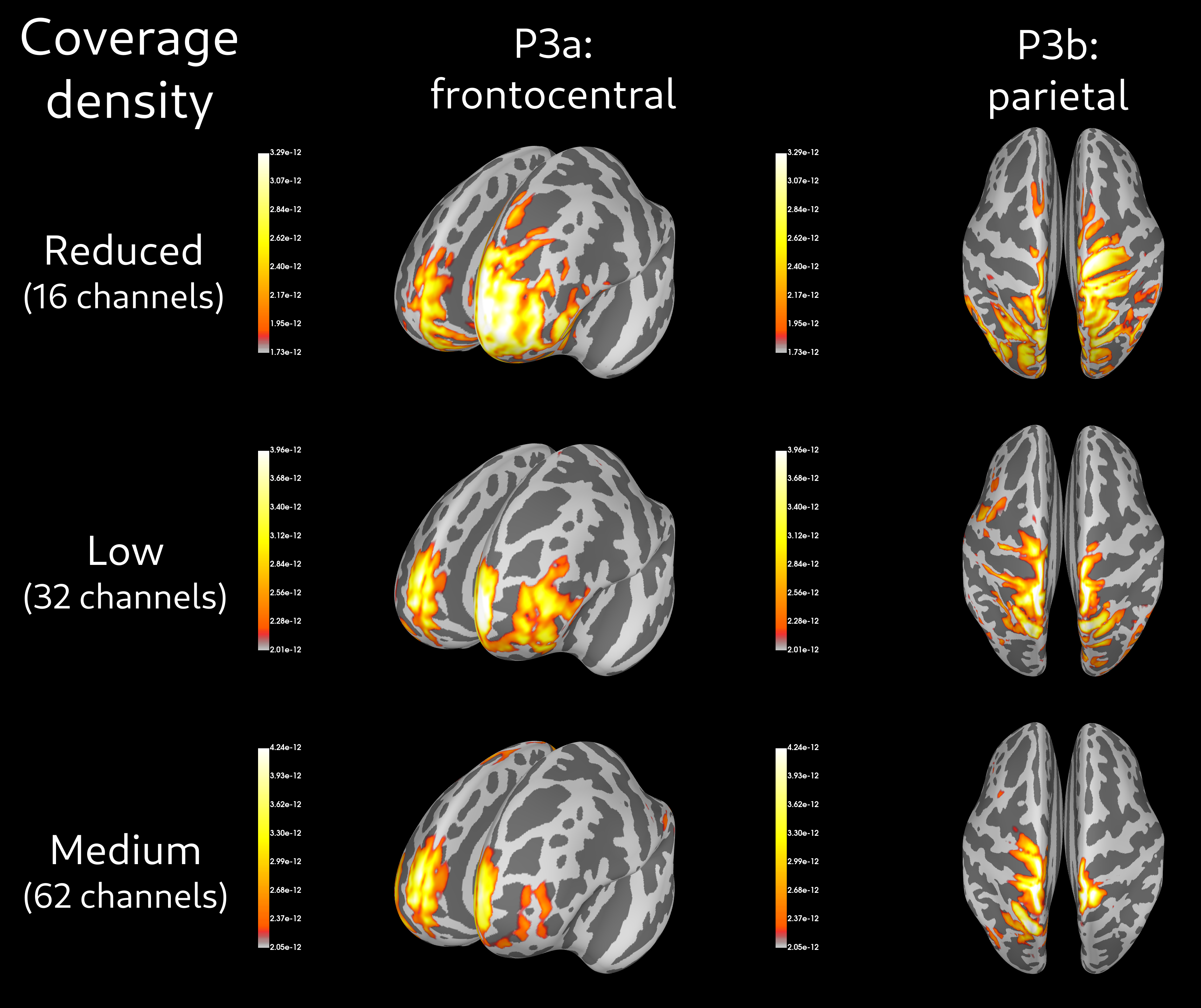}
	\caption
	[Channel configuration-dependent source reconstruction]
	{Source localization accuracy across electrode densities (62, 32, and 16 channels) during P300 for a representative subject: left displays P3a and right P3b. Color bars indicate normalized current density ($\mu A/mm^2$).}
	\label{fig:Blurring}
\end{figure}

The inverse solution is computed at each time point using regularized reconstruction. Figure \ref{fig:Blurring} compares results across channel configurations for a representative subject corresponding to the P3a and P3b subcomponents. The medium-density (62-channel) configuration successfully resolves the characteristic central activation of the P3a, while this focal pattern becomes indistinguishable in both low-density (32-channel) and reduced (16-channel) setups. As anticipated, progressive spatial blurring accompanies electrode reduction, particularly evident in the degraded frontal P3a and parietal P3b topographies, demonstrating how channel count systematically affects spatial localization accuracy.

\subsection{Cortical Network Connectivity Analysis}
\begin{figure}[!htp]
	\centering
	\includegraphics[width=8cm]{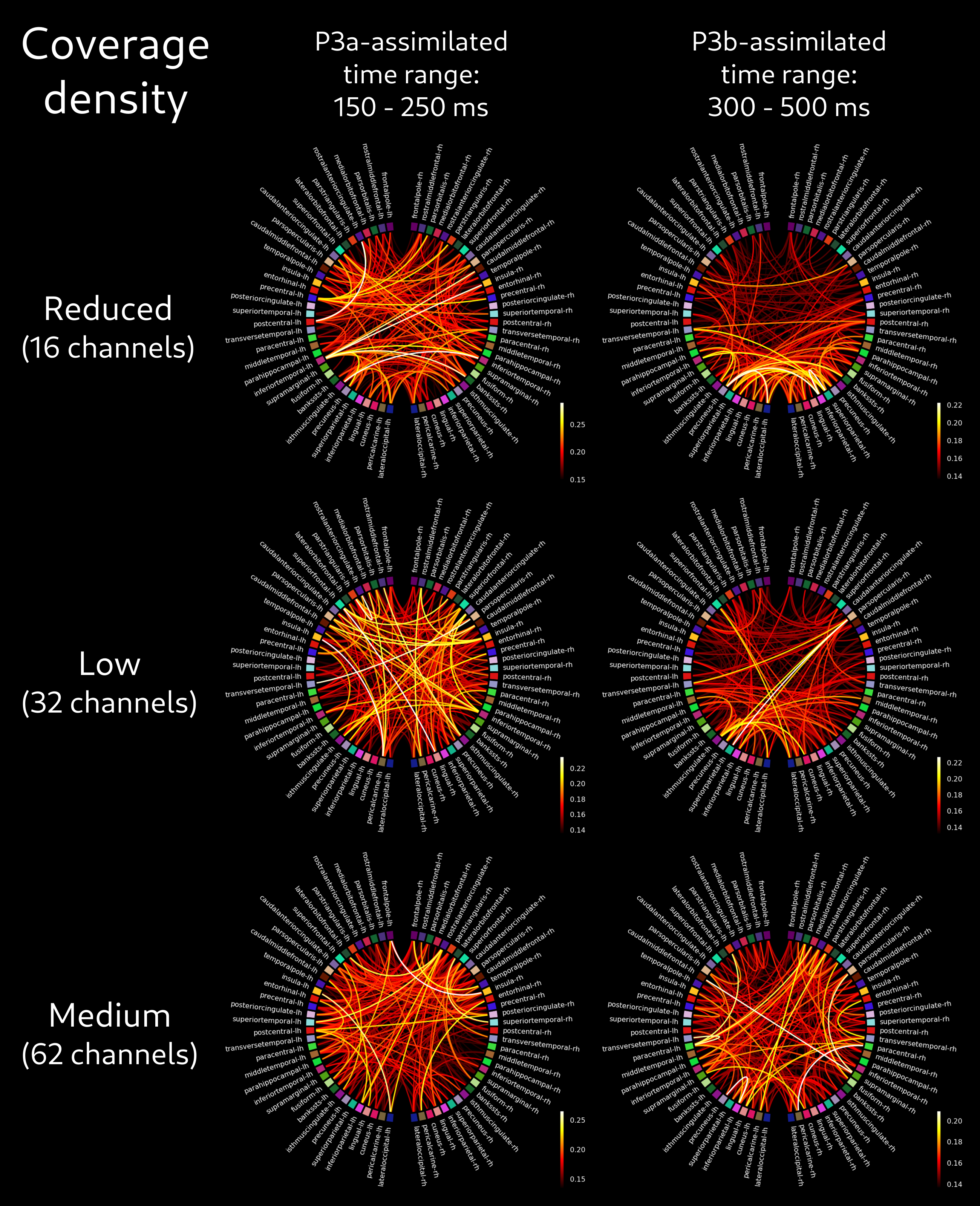}
	\caption
	[P300 subcomponent functional connectivity]
	{Circular graphs display the strongest connections between FreeSurfer parcels during 150 -- 250 ms (representing P3a) and 300 -- 500 ms (for P3b). Paler edge colors indicate stronger correlation values.}
	\label{fig:Connectivity}
\end{figure}

The reconstructed source activity undergoes spatial averaging within each of the 75 cortical regions defined by the FreeSurfer parcellation scheme. This regional aggregation transforms the voxel-level source estimates into anatomically meaningful parcels while reducing computational complexity. The resulting time-series data form the basis for connectivity analysis, visualized through circular graphs organized by axial plane positioning, as illustrated in Figure \ref{fig:Connectivity}. 

Component-specific temporal windows maximize the isolation of P300 subcomponents. The P3a analysis focuses on the 150-250 ms post-stimulus interval, capturing its characteristic frontocentral activation during the ascending phase while minimizing P3b contamination. The P3b assessment employs a 300-500 ms window to accommodate its broader temporal distribution and lower amplitude, with the extended duration compensating for potential residual P3a overlap. This temporal partitioning aligns with established P300 dynamics while optimizing the detection of distinct subnetworks.

Phase Lag Index (PLI) analysis characterizes transient network dynamics during P300 subcomponent windows (P3a: 150-250 ms; P3b: 300-500 ms), effectively rejecting volume conduction artifacts by excluding zero-lag phase interactions \cite{Stam_2007}. To establish robust regions of interest (ROIs), analysis keeps only the top 20 connection pairs per session, with final ROIs defined as the three brain regions exhibiting highest cross-participant consistency. Single-trial classification detects activation maxima within these ROIs (-200 to +700 ms), with P300-positive trials defined by peaks occurring during the neurophysiological response window (+200 to +550 ms), leveraging eLORETA's anatomical specificity.

\begin{figure}[!htp]
	\centering
	\includegraphics[width=8cm]{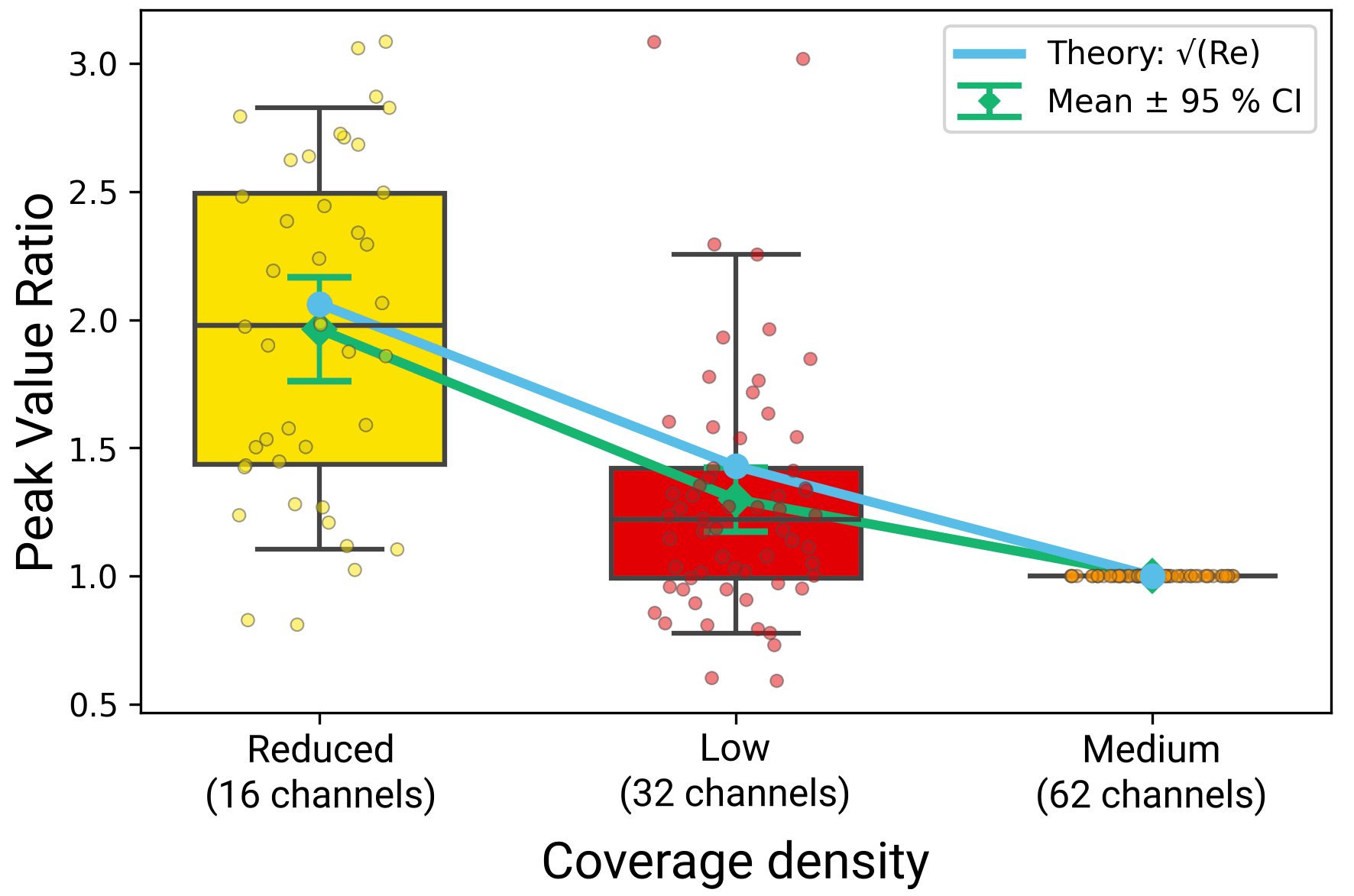}
	\caption
	[Peak Val Ratio for P3a]
	{Comparison of experimental and a theoretically approximated Peak Value Ratios for the P3a component (150–250 ms window).} 
	\label{fig:PeakValRatio}
\end{figure}

For each detected P300 response across the three electrode configurations, the corresponding maximum activation value from the medium-coverage configuration served as the normalization reference. Figure \ref{fig:PeakValRatio} displays the Peak Value Ratios across all single sessions, excluding statistical outliers defined as values beyond ±3 standard deviations from the mean. Notably, the P3b component shows the same robust scaling pattern, further validating these findings across P300 subcomponents (not represented here for visual clarity). This normalization approach facilitates direct comparison of signal fidelity degradation across spatial sampling densities while controlling for inter-session variability.

{In this study, localization accuracy is quantified using peak P300 activation values based on three key properties: (1) the component's robust stimulus-locked morphology across sessions, (2) the maximal signal-to-noise ratio at peak activation representing optimal spatial precision, and (3) its direct correlation with BCI classification performance through detectable activation strength. The observed localization accuracy follows a square root relationship with electrode density, where reducing electrode count by a factor $R_e$ decreases accuracy by $\sqrt{R_e}$. This theoretical approximation holds within statistical bounds, as demonstrated by the 95\% confidence intervals of the experimental data encompassing the predicted $\sqrt{R_e}$ scaling. }


\section{Discussion}

\subsection{Methodological Trade-offs and Validation}
The standardized pipeline employed in this study isolates electrode density as the primary variable affecting source reconstruction accuracy, revealing fundamental limitations of sparse montages. While smaller configurations (16-32 channels) preserve superficial P300 topography, they obscure deeper generators and induce spatiotemporal distortions through volume conduction effects. These findings validate theoretical predictions about spatial sampling requirements, demonstrating that even robust methods like eLORETA cannot fully compensate for inadequate electrode coverage. The derived $\sqrt{R_e}$ scaling law represents an attempt to a generalizable framework applicable to any BCI paradigm and signal of interest, with its intentionally coarse approximation abstracting paradigm-specific factors to reveal fundamental spatial sampling constraints. The use of template head models and conservative channel exclusion criteria, though necessary for cross-subject consistency, likely provides a conservative estimate of achievable resolution — particularly for patient populations where individual anatomical variability may exacerbate localization errors. The P300 paradigm serves as an ideal test case here, its well-characterized properties mitigating inter-subject variability while challenging spatial resolution across cortical and subcortical networks.

\subsection{Clinical Translation and Implementation Strategies}
From a practical perspective, our results advocate for a tiered approach to electrode selection based on clinical objectives. Higher density arrays are likely to prove essential for diagnostics targeting deep structures, whereas simplified montages may suffice for basic P300 detection in communication BCIs. Practical barriers to dense array implementation — particularly extended setup times for compromised populations — could be addressed through adaptive protocols, e.g., rapid application systems, or immersive patient engagement during preparation. Crucially, the observed temporal compression in low-density configurations warns against overinterpreting functional connectivity metrics, as sparse sampling artificially synchronizes superficially reconstructed sources. These insights directly inform BCI design trade-offs, suggesting that optimal configurations must balance spatial precision against user-specific constraints.

\subsection{Future Directions for Precision Source Imaging}

Future research should address three key limitations of the current work. First, subject-specific head modeling would disentangle true anatomical variability from template-induced blurring, particularly for patient populations with structural abnormalities. Second, dynamic connectivity analysis using individualized P300 latency windows — derived from single-trial estimates rather than fixed ranges — could enhance sensitivity to network reorganization. Finally, combining high-density EEG with auxiliary modalities (e.g., fNIRS for hemodynamic correlates or MEG for complementary spatial resolution) may enable surrogate markers of deep source activity even when using practical electrode counts. This multimodal approach would circumvent the fundamental physics limitations of scalp EEG while maintaining clinical feasibility, potentially achieving resolution comparable to dense arrays through data fusion rather than channel proliferation.

\section{Conclusion}

This work establishes a methodological framework for evaluating EEG electrode configurations through quantitative metrics of spatial localization accuracy and source connectivity -- key factors influencing BCI performance. As a first systematic assessment of density effects using standardized protocols, we demonstrate that 62-channel configurations reliably reconstruct both superficial and deep P300 generators, while sparser montages show progressive degradation and introduce artificial temporal synchronization of cortical activity.

Our findings directly inform experimental design by quantifying the $\sqrt{R_e}$ scaling relationship between electrode count and localization error. This allows researchers to select the minimal electrode density required based on (1) the acceptable margin of error for their target neural phenomena or (2) the expected effect size of interest. While these findings require systematic validation across diverse populations and tasks, the consistent effects observed across both spatial and temporal domains underscore the value of this dual-metric approach as a foundation for optimizing practical EEG systems.



\begin{thebibliography}{00}
\bibitem{Grover_2016} P. Grover, ``Fundamental limits on source-localization accuracy of EEG-based neural sensing,'' in 2016 IEEE International Symposium on Information Theory (ISIT), pp. 1794 -- 1798, Jul. 2016. 

\bibitem{Chaudhary_2016} U. Chaudhary, N. Birbaumer, and A. Ramos-Murguialday,  ``Brain–computer interfaces for communication and rehabilitation,'' in Nature Reviews Neurology, vol. 12, pp. 513--525, Aug. 2016. 


\bibitem{Hauk_2022} O. Hauk, M. Stenroos, and M. S. Treder, ``Towards an objective evaluation of EEG/MEG source estimation methods – The linear approach,,'' in Neuroimage, vol. 255, pp. 119177, Jul. 2022.


\bibitem{Grova_2008} C. Grova, J. Daunizeau, E. Kobayashi, A. P. Bagshaw, J. M. Lina, F. Dubeau, and J. Gotman, ``Concordance between distributed EEG source localization and simultaneous EEG-fMRI studies of epileptic spikes,'' in Neuroimage, vol. 39 (2), pp. 755 -- 774, Jan. 2008.

\bibitem{Sohrabpour_2015} A. Sohrabpour, Y. Lu, P. Kankirawatana, J. Blount, H. Kim, and B. He, ``Effect of EEG electrode number on epileptic source localization in pediatric patients,'' in Clinical Neurophysiology, vol. 126 (3), pp. 472 -- 480, Jul. 2015.

\bibitem{AkalinAcar_2013} Z. Akalin Acar and S. Makeig, ``Effects of forward model errors on EEG source localization,'' in Brain topography, vol. 26, pp. 378 -- 396, Jan. 2013.

\bibitem{Unnwongse_2023} K. Unnwongse, S. Rampp, T. Wehner, A. Kowoll, Y. Parpaley, M. von Lehe, B. Lanfer, M. Rusiniak, C. Wolters and J. Wellmer, ``Validating EEG source imaging using intracranial electrical stimulation,'' in Brain Communications, vol. 5 (1), fcad023, Feb. 2023.

\bibitem{Brodbeck_2011} V. Brodbeck, L. Spinelli, A. M. Lascano, M. Wissmeier, M. I. Vargas, S. Vulliemoz, C. Pollo, K. Schaller, C. M. Michel and M. Seeck, ``Electroencephalographic source imaging: a prospective study of 152 operated epileptic patients,'' in Brain, vol. 134 (10), pp. 2887 -- 2897, Aug. 2011.

\bibitem{Eom_2022} T. H. Eom, ``Electroencephalography source localization,'' in Clinical and Experimental Pediatrics, vol. 66 (5), pp. 201, Nov. 2022.

\bibitem{TrujilloBarreto_2004} N. J. Trujillo-Barreto, E. Aubert-Vázquez, and P. A. Valdés-Sosa, ``Bayesian model averaging in EEG/MEG imaging,'' in NeuroImage, vol. 21 (4), pp. 1300 -- 1319, 2004.

\bibitem{Michel_2019} C. M. Michel, and D. Brunet, ``EEG source imaging: a practical review of the analysis steps,'' in Frontiers in neurology, vol. 10 (325), Apr. 2019.

\bibitem{Dattola_2020} S. Dattola, F. La Foresta, L. Bonanno, S. De Salvo, N. Mammone, S. Marino and F. C. Morabito, ``Effect of sensor density on eLORETA source localization accuracy,'' in Neural Approaches to Dynamics of Signal Exchanges, pp. 403 -- 414, 2020.

\bibitem{PascualMarqui_2018} R. D. Pascual-Marqui, P. Faber, T. Kinoshita, K. Kochi, P. Milz, K. Nishida and M. Yoshimura, ``Comparing EEG/MEG neuroimaging methods based on localization error, false positive activity, and false positive connectivity,'' in BioRxiv, pp. 269753, Feb. 2018.

\bibitem{Hedges_2016} D. Hedges, R. Janis, S. Mickelson, C. Keith, D. Bennett, and B. L. Brown, ``P300 Amplitude in Alzheimer’s Disease: A Meta-Analysis and Meta-Regression,''	in Clinical EEG and Neuroscience, vol. 47, no. 1, pp. 48--55, Sep. 2014. 

\bibitem{eye_bci_multi_dataset} E. Guttmann-Flury, X. Sheng and X. Zhu, ``Eye-BCI multimodal dataset,'' Synapse 2024. Available: https://doi.org/10.7303/syn64005218 

\bibitem{GuttmannFlury_Dataset_2025} E. Guttmann-Flury, X. Sheng and X. Zhu, ``Dataset combining EEG, eye-tracking, and high-speed video for ocular activity analysis across BCI paradigms,'' in Sci. Data, vol. 12 (587), Apr. 2025. 

\bibitem{GuttmannFlury_APriori_2019} E. Guttmann-Flury, X. Sheng and X. Zhu, ``A priori sample size determination for the number of subjects in an EEG experiment,'' in 41st Annual International Conference of the IEEE Engineering in Medicine and Biology Conference (EMBC), Berlin, Germany, pp. 5180--5183, Jul. 2019.	

\bibitem{Koessler_2009} L. Koessler, L. Maillard, A. Benhadid, J. P. Vignal, J. Felblinger, H. Vespignani and M. Braun, ``Automated cortical projection of EEG sensors: anatomical correlation via the international 10–10 system,'' in Neuroimage, vol. 46 (1), pp. 64--72, Feb. 2009.	

\bibitem{HerculanoHouzel_2009} S. Herculano-Houzel, ``The human brain in numbers: a linearly scaled-up primate brain,'' in Frontiers in human neuroscience, vol. 3 (31), pp.857, Nov. 2009.	

\bibitem{Nemaire_2023} M. Nemaire, ``Inverse potential problems, with applications to quasi-static electromagnetics,'' in General Mathematics [math.GM]. Université de Bordeaux, 2023. English, Apr. 2023.	

\bibitem{Fischl_1999} B. Fischl, M. I. Sereno, R. B. Tootell and A. M. Dale, ``High‐resolution intersubject averaging and a coordinate system for the cortical surface,'' in Human brain mapping, vol. 8 (4), pp.272 -- 284, Jul. 1999.	

\bibitem{Gramfort_2013} A. Gramfort, M. Luessi, E. Larson, D. A. Engemann, D. Strohmeier, C. Brodbeck, R. Goj, M. Jas, T. Brooks, L. Parkkonen and M. S. Hämäläinen, ``MEG and EEG data analysis with MNE-Python,'' in Frontiers in Neuroscience, vol. 7 (267), pp.1 -- 13, Dec. 2013.	

\bibitem{GuttmannFlury_BadChannel_2025} E. Guttmann-Flury, W. Wei and S. Zhao, ``Automatic Blink-based Bad EEG channels Detection for BCI Applications,'' in 47th Annual International Conference of the IEEE Engineering in Medicine and Biology Conference (EMBC), Copenhagen, Denmark, Jul. 2025.	

\bibitem{Kayser_2015} J. Kayser and C. E. Tenke, ``On the benefits of using surface Laplacian (current source density) methodology in electrophysiology,'' in International Journal of Psychophysiology, vol. 97, no. 3, pp. 171, Sep. 2015. 

\bibitem{Stam_2007} C. J. Stam, G. Nolte, and A. Daffertshofer, ``Phase lag index: assessment of functional connectivity from multi channel EEG and MEG with diminished bias from common sources,'' in Human Brain Mapping, vol. 28, no. 11, pp. 1178 -- 1193, Jan. 2007.	


\end{thebibliography}
\end{document}